\documentclass[aps,prd,preprintnumbers,showpacs,nofootinbib,twocolumn]{revtex4}
\usepackage{graphicx}
\usepackage{bm}
\usepackage{amsmath}
\usepackage{amssymb}
\usepackage{amsthm}

\newcommand{\sigl}{\sigma_\lambda}
\newcommand{\lsim}   {\mathrel{\mathop{\kern 0pt \rlap
  {\raise.2ex\hbox{$<$}}}
  \lower.9ex\hbox{\kern-.190em $\sim$}}}
\newcommand{\gsim}   {\mathrel{\mathop{\kern 0pt \rlap
  {\raise.2ex\hbox{$>$}}}
  \lower.9ex\hbox{\kern-.190em $\sim$}}}

\begin{document}

\title{Dark energy, non-minimal couplings and the origin of cosmic magnetic fields}

\author{Jose Beltr\'an Jim\'enez}
\author{Antonio L. Maroto}

\affiliation{Departamento de F\'isica Te\'orica, Universidad Complutense 
de Madrid, 28040, Madrid, Spain.}


\date{\today}

\begin{abstract}
In this work we consider the most general 
electromagnetic theory in curved space-time 
leading to linear second order differential equations,
including non-minimal
couplings to the space-time curvature.  We assume 
the presence
of a temporal electromagnetic background whose energy density plays the role
of dark energy, as has been recently suggested. Imposing the consistency
 of the theory  in the weak-field limit, we show that it 
reduces to 
standard electromagnetism in the presence of an effective electromagnetic
current which is generated by the momentum density of the matter/energy 
distribution, even for neutral sources. This implies that in the presence
of dark energy, the motion of large-scale structures 
generates magnetic fields. Estimates of the present 
amplitude of the generated seed fields for typical spiral 
galaxies could reach $10^{-9}$ G without any amplification. 
In the case of compact rotating
objects, the theory predicts their magnetic moments  
to be  related to their 
angular momenta in the way suggested by
the so called Schuster-Blackett conjecture.

\end{abstract}

\maketitle
\section{Introduction}
The origin of the magnetic fields observed in galaxies and galaxy clusters
with large coherence lengths and strengths around $10^{-6}$ G  still remains
an open problem in astrophysics  \cite{Widrow} (recent works 
\cite{extragalactic} also show  
evidence for the existence of extragalactic magnetic fields with 
strengths above $3\times 10^{-16}$ G).
Two different types of mechanisms
have been proposed for the generation of such fields. 
On one hand, we  have 
the primordial field hypothesis, i.e.   
relic fields from the early universe 
with comoving strengths around $10^{-10}-10^{-12}$ G 
 are amplified to the present values 
in the protogalactic collapse. On the other, much
weaker fields around $10^{-19}$ G at decoupling time 
could have been amplified by the galactic rotation through 
a dynamo mechanism. 
In both cases, preexisting seed fields are required.
In fact, 
there are also several  proposals for the generation
of fields which could seed a galactic dynamo. They include 
astrophysical mechanisms \cite{Harrison}, 
production during inflation \cite{Turner}, in phase transitions \cite{QLS},
by spontaneous breaking of Lorentz invariance \cite{Mota} 
or by metric perturbations \cite{pert}. Nevertheless, it 
has been argued that the timescales for 
dynamo amplification may be too long to explain the observed fields in 
young objects \cite{Widrow}. In addition,  the origin of the 
stronger  large-scale seeds in the primordial approach 
is even more problematic. 

A very interesting framework for magnetic field generation is 
the possibility that the standard electromagnetic theory  
could be modified in the presence of gravity. 
Thus in \cite{Turner}, couplings of the electromagnetic field
to the space-time curvature were proposed as a way of 
producing magnetic fields during inflation. In  
this paper we will consider a generalized electromagnetic action in 
curved space-time, including also non-minimal couplings.  
The crucial difference with respect to previous works  
is that we allow for the presence of a homogeneous 
temporal 
electromagnetic background potential. 
This is motivated by the fact that, as has been recently shown 
\cite{EM}, the presence of   
temporal electromagnetic potentials on cosmological scales 
could play the role of 
dark energy. Indeed, this type of fields can be amplified from quantum 
fluctuations during inflation in a completely analogous way to 
metric perturbations. The initial amplitude of the field being given by  
$\langle A_0^2\rangle^{1/2}\sim H_I$, where $H_I$ 
is the value of the Hubble parameter during inflation.
The field is then shown to grow linearly in time in the matter and
radiation eras, the corresponding energy density  
on cosmological scales  behaving as 
a cosmological constant. Interestingly, the predicted value
of the cosmological constant  agrees with observations provided 
 inflation took place at the electroweak scale. In such a case,
the present value of the background field would 
be $\bar A_0\simeq 0.3 \,M_P$.
Here we  show that the non-minimal coupling of 
the temporal background to the 
space-time curvature implies that  
the energy-momentum 
density of any matter/energy distribution 
generates an effective electromagnetic current, even for neutral sources. 
This allows to establish a natural link  between 
 dark energy and the origin of cosmic magnetic fields. 

\section{Generalized electromagnetism}

Let us consider the most general expression for the electromagnetic action
in the presence of gravity, including all the possible terms leading to linear
second order differential equations:
\begin{eqnarray}
S=\int d^4x\sqrt{-g}\left[-\frac{1}{4}F_{\mu\nu}F^{\mu\nu}
+\frac{\lambda}{2}(\nabla_\mu A^\mu)^2\right.\nonumber\\ \left.
+\sigma R_{\mu\nu}A^\mu A^\nu+\omega R A_\mu A^\mu\right].
\label{action}
\end{eqnarray}
Notice that this expression does not contain any dimensional
parameter or potential term. 
The minimal case with $\sigma=\omega=0$ was studied in detail
in \cite{EM} and the possibility of generating cosmic magnetic fields 
in this case has been considered recently in \cite{minimal}. 
In this action, the $\lambda$ parameter  can be fixed 
by choosing a normalization of the non-transverse modes and 
$\sigma$ and $\omega$ are arbitrary dimensionless constants. 
In order to fix them, we will consider the weak-field limit
of the theory. Thus,  the space-time metric can be written as
a small 
perturbation around  Minkowski space-time, $g_{\mu\nu}=
\eta_{\mu\nu}+h_{\mu\nu}$ and the electromagnetic potential reads 
 $A_\mu=\bar{A}_\mu+a_\mu$ with 
$\bar{A}_\mu=\bar{A}_0\delta^0_\mu$ and $\bar{A}_0$ constant. 
The background electromagnetic field is determined by the 
corresponding cosmological value and therefore it could evolve on 
cosmological timescales. However, for local experiments it is a
good approximation to assume it 
constant (in agreement with the flat space-time 
background). Notice that the electric and magnetic fields associated to 
$\bar{A}_\mu$ identically vanish. 
The corresponding Maxwell equations obtained from (\ref{action}) read
to first order:
\begin{eqnarray}
\partial_\nu F^{\mu\nu}+\lambda\partial^\mu(\nabla_\nu A^\nu)_{(1)}=
J_{g}^\mu.
\label{Maxwell1}
\end{eqnarray}
where $F_{\mu\nu}=\partial_\mu a_\nu-\partial_\nu a_\mu$, 
 $(\nabla_\nu A^\nu)_{(1)}$ denotes the contribution to
first order and the non-minimal terms  give rise to
 an effective current given also to first order 
by: $J_{g}^\mu=2(\sigma\, R^{\mu \nu}_{(1)}+\omega\, 
R_{(1)}\eta^{\mu \nu})\bar{A}_\nu$.  
 Imposing  this effective
 current to be conserved, i.e. $\partial_\mu J^\mu_{g}=0$, 
we obtain $\sigma=-2\omega$, i.e.
the non-minimal coupling must involve the conserved Einstein 
tensor $G_{\mu\nu}=R_{\mu\nu}-\frac{1}{2}Rg_{\mu\nu}$. 
Notice also that conservation 
implies that taking the divergence of (\ref{Maxwell1}) we get
$\Box (\nabla_\mu A^\mu)_{(1)}=0$, i.e. to first order 
it is possible to impose the
Lorenz condition $\nabla_\mu A^\mu=0$ at the classical level as
in ordinary electromagnetism, so that the 
$\lambda$ term disappears. Thus, for weak gravitational fields
we recover ordinary Maxwell electromagnetism, the only difference
is the appearance of a gravitationally-generated  
electromagnetic current. Notice that this current is only present
provided the background electromagnetic potential is non-vanishing and
in the presence of space-time curvature.

According to the previous discussion, the form of the action 
will be given by:
\begin{eqnarray}
S&=&\int d^4x\sqrt{-g}\left[-\frac{1}{4}F_{\mu\nu}F^{\mu\nu}\right.\nonumber \\
&+&\left.\frac{\lambda}{2}(\nabla_\mu A^\mu)^2
+\sigma G_{\mu\nu}A^\mu A^\nu\right]\label{action}
\end{eqnarray}

\section{Consistency and stability}

This theory is a particular 
case of the more general class of vector-tensor theories \cite{Will}. 
These 
theories usually give rise to modifications of the gravitational 
interaction at small (Solar System) scales which are encoded in 
the corresponding PPN parameters. For this particular case, the 
PPN parameters are: $\gamma-1\simeq-16\pi G\sigma A_{\odot}^2$, 
$\beta-1\simeq-16\pi G\sigma A_{\odot}^2$, $\alpha_1\simeq
-64\pi G\sigma A_{\odot}^2$, $\alpha_2\simeq-16\pi G\sigma A_{\odot}^2$, 
where $A_{\odot}$ is the background amplitude at Solar System scales and we 
have assumed $\vert\sigma\vert\ll 1$, keeping only the leading 
order in the expansion. The most stringent constraint 
on the PPN parameters is $\vert\gamma-1\vert\leq 2.3\times 10^{-5}$ ,
 which imposes the corresponding limit on $\vert \sigma A_{\odot}^2\vert$. 
If we assume that the amplitude of the electromagnetic field at 
Solar System scales resembles the cosmological value 
$A_{\odot}\simeq \bar{A}_0\simeq 0.3 M_P$, 
we obtain the constraint $\vert\sigma\vert \lsim 10^{-5}$.

Let us now study the stability of the theory by analyzing 
the behavior of the inhomogeneous perturbations around the 
Minkowski background. As usual, we shall decompose both the electromagnetic 
perturbation $a_\mu$ and the metric perturbation $h_{\mu\nu}$ in
Fourier modes and separate them into scalar, vector 
and tensor contributions (we follow the same procedure as in 
\cite{VT}). The corresponding propagation speeds for 
the perturbations are:
\begin{eqnarray}
c_s^2&=&1\\
c_v^2&=&\frac{1-8\pi G \sigma(1-2\sigma)\bar A_0^2}{1-8\pi G\sigma \bar A_0^2}
\simeq1+16\pi G\sigma^2 \bar A_0^2\\
c_t^2&=&\frac{1+8\pi G\sigma \bar A_0^2}{1-8\pi G\sigma \bar A_0^2}\simeq 1+16\pi G\sigma 
\bar A_0^2
\end{eqnarray}
where we have expanded for $\vert\sigma\vert \ll 1$. 
Notice that the scalar modes propagate at the speed of light 
irrespective of the value of the parameter $\sigma$. 
However,  the speed of photons $c_v$ 
would be larger than the "speed of light" $c=1$ 
which determines the null cones 
of the Minkowski geometry. This in principle could give rise to
inconsistencies with causality in the theory.
However, it is known that in scenarios with 
violations of the strong equivalence principle, 
as the one considered here, superluminal propagation can be consistent
with causality, provided {\it stable causality} is ensured \cite{HE}. 
For that purpose,
if the new light cone can be written as ${\cal G}^{\mu\nu}k_\mu k_\nu=0$, 
then there must exist a globally defined function $f$, 
such that $\nabla_\mu f$ must
be non-vanishing and timelike  everywhere 
with respect to $({\cal G}^{-1})_{\mu\nu}$. In our case, the light cones for vectors  can
be written as:
\begin{equation}
\left[\left(1+16\pi G\kappa \bar A^2\right)\eta_{\mu\nu}
-16\pi G\kappa\bar{A}_\mu\bar{A}_\nu\right]k^\mu k^\nu=0
\end{equation}
with $\kappa=\sigma^2/(1-\sigma A^2)$, 
whereas for tensors:
\begin{equation}
\left[\left(1+8\pi G\sigma \bar A^2\right)\eta_{\mu\nu}
-16\pi G\sigma\bar{A}_\mu\bar{A}_\nu\right]k^\mu k^\nu=0
\end{equation}
Since $\bar A_0\lsim M_P$ and $\vert\sigma\vert\ll 1$,  
in both cases, the effective metric 
$({\cal G}^{-1})_{\mu\nu}$ is a small perturbation 
with respect to Minkowski.   
This implies that we can use the  time coordinate $t$ as the globally defined
function $f$. Thus, we see that, 
for small $\sigma$, the theory does not exhibit classical instabilities or 
causality inconsistencies.  

In order to study the presence of quantum instabilities (ghosts), 
we analyze 
the positiveness of the energy density of the three types of
perturbations considered before. Thus, 
we define the energy for the modes as
\cite{Eling,VT}:
\begin{equation}
\rho=\left<T_{00}^{(2)}-\frac{1}{8\pi
G}G_{00}^{(2)}\right>\label{defe}
\end{equation}
where $T_{\mu\nu}^{(2)}$ and $G_{\mu\nu}^{(2)}$ are the
energy-momentum tensor of the vector field and the Einstein's
tensor calculated up to quadratic terms in the perturbations and
$\left<\cdots\right>$ denotes an average over spatial regions.
Although the calculation has been performed in the longitudinal gauge,
both, mode frequencies and energies, do not depend on the gauge 
choice.

For scalar modes we find 
that the energy density vanishes identically if we impose the Lorenz 
condition, as in ordinary electromagnetism (see \cite{EM} for expanding
backgrounds).  For vector and tensor modes, the energy densities are:
\begin{eqnarray}
\rho_v&=&2k^2\frac{1-8\pi G\sigma\left[2+8\pi G\sigma(2\sigma-1)
\bar A_0^2\right]\bar A_0^2}
{(1-8\pi G\sigma \bar A_0^2)^2}\vert \vec{C}\vert ^2\nonumber\\
&\simeq& 2k^2(1-16\pi G\sigma^3 \bar A_0^4)
\vert \vec{C}\vert ^2\\
\rho_t&=&k^2\frac{1-8\pi G\sigma(2+8\pi G\sigma \bar A_0^2)\bar A_0^2}{1-8\pi G\sigma 
\bar A_0^2}
\left(\vert C_{\oplus}
\vert^2 +\vert C_{\otimes} \vert ^2\right)\nonumber\\
&\simeq& k^2(1-8\pi G\sigma \bar A_0^2)\left(\vert C_{\oplus}
\vert^2 +\vert C_{\otimes} \vert ^2\right)
\end{eqnarray}
where $\vec{C}$ is the amplitude of the Fourier mode for the vector 
modes and $C_{\oplus, \otimes}$ are the amplitudes of the two 
polarizations of the gravitational waves. From these expressions 
we see that the theory is also free from quantum instabilities 
for small $\vert\sigma\vert$.

\section{Cosmological evolution}

In the following we shall show that, 
due to the smallness of the parameter $\sigma$, 
the cosmological evolution of the homogeneous mode 
becomes modified in a negligible way by the
 presence of the coupling to the Einstein tensor. 
This ensures that the  
inflationary generation and evolution discussed in \cite{EM} is also a 
good description in the non-minimal case.
We shall consider
 an electromagnetic field of the form $A_\mu=(A_0(t),0,0,A_z(t))$ in
 a FLRW metric $ds^2=dt^2-a(t)^2d\vec{x}^2$. In this case, the equations 
of motion read:
\begin{eqnarray}
\ddot{A}_0+3 H\dot{A}_0+3\left(\dot{H}-2\sigl H^2\right)A_0=0\\
\ddot{A}_z+H\dot{A}_z+\sigma(4\dot{H}+6H^2)A_z=0
\end{eqnarray}
where $\sigl=\frac{\sigma}{\lambda}$. 
In a de-Sitter inflationary era with $H=H_I$ constant, the growing
mode solutions behave for small $\sigma$ as:
\begin{eqnarray}
A_0(t)\propto \exp({2\sigl H_It}), \;\;\; 
A_z(t)\propto \exp({-6\sigma H_It}) 
\end{eqnarray}
During the 
radiation and matter dominated epochs in which $H=p/t$ with $p=1/2$ 
and $p=2/3$ respectively, the solutions are:
\begin{eqnarray} 
A_0(t)\propto t^{1+3\sigl/5},\;\;\;A_z(t)\propto t^{1/2
+\sigma}
\end{eqnarray}
in the radiation era, and 
\begin{eqnarray}
A_0(t)\propto t^{1+8\sigl/9}, \;\;\; 
A_z(t)\propto t^{1/3}
\end{eqnarray}
 in the matter era. We see that  the the only effect of the 
non-minimal coupling is a slight modification in the power
exponents.
Finally, in a universe dominated by the electric potential $A_0(t)$,  
we have a power law expansion of the form $a(t)
\propto t^{-\frac{\lambda}{2\sigma}}$. For small 
$\sigma$, we have an accelerated 
expansion which corresponds to a quasi 
de Sitter phase with slow-roll parameter $\epsilon=-\frac{2\sigma}{\lambda}$. 
Notice that, in the limit $\sigma\rightarrow 0$, we also recover the pure de 
Sitter solution found in the minimal case. 

\section{Effective electromagnetic current: gravitational magnetism}

Let us now consider the possible effects of the new 
effective electromagnetic 
current  $J_{g}^\mu=2\sigma G^{\mu 0}\bar A_0$ in (\ref{Maxwell1}). 
Using 
Einstein equations to relate $G^{\mu\nu}$ to the matter content, we 
obtain:
\begin{eqnarray}
J_{g}^\mu=16\pi G\sigma T^{\mu 0}\bar A_0
\end{eqnarray}
so that the effective 
electromagnetic current is essentially determined by the four-momentum 
density.  Moreover, if we assume $T^{\mu\nu}=
(\rho+p)u^\mu u^\nu-p\eta^{\mu\nu}$ at first order, 
we can see that the energy density
of any perfect fluid 
has an associated electric charge density given, for small velocities, 
by: 
\begin{equation}
\rho_g=J_g^0=16\pi G\sigma\rho \bar A_0\label{effectiveq}
\end{equation}
and the three-momentum density generates an electric current density 
given by 
\begin{equation}
\vec{J}_g=16\pi G\sigma (\rho+p)\vec{v}\bar A_0
\end{equation}

This theory effectively realizes the old conjecture by Schuster, Einstein  
and Blackett \cite{old} 
of gravitational magnetism, i.e. neutral mass currents generating
electromagnetic fields. Early attempts to encompass this conjecture in
a gravitational theory can be found in \cite{early}. 

In the case of a particle of mass $m$ at rest, 
(\ref{effectiveq}) introduces a small contribution to the 
{\it active} electric charge 
(the source of the electromagnetic field), given by
$\Delta q= 16\pi G\sigma m \bar A_0\simeq 15\sigma (m/M_P)$, 
but does not modify the 
{\it passive} electric charge (that determining the coupling 
to the electromagnetic field).  
In fact, this would give different active charges to electrons and 
protons due to their mass difference and,  in addition, would provide the 
neutron with a non-vanishing active electric charge. However, the effect is 
very small in both cases $\Delta q\simeq 4\sigma  10^{-18}e$
where  $e=0.303$ is the 
electron charge 
in Heaviside-Lorentz units. Present limits on the 
electron-proton charge asymmetry  and neutron charge are both of the order
$10^{-21}e$ \cite{limits}, 
implying $\vert \sigma\vert\lsim 10^{-3}$ which is less stringent than
the PPN limit discussed before. Notice also that photons would acquire a 
non-vanishing active electric charge. However tight existing limits  
imposed by deflection of radio pulsar emission by galactic 
magnetic fields \cite{Raffelt}
only apply to passive charge which is not modified gravitationally.


On the other hand, for any compact object, even in the case 
it is neutral, the effective electric current will  
generate an intrinsic magnetic moment $\vec{m}=\frac{1}{2}
\int\vec{r}\times\vec{J}_{g}(\vec{r})d^3\vec{r}$ given by:
\begin{equation}
\vec{m}=\beta\frac{\sqrt{G}}{2}\vec{L}\label{S-Blaw}
\end{equation}
with $\vec L$ the corresponding angular momentum and 
$\beta$ a constant parameter whose value is:
\begin{equation}
\beta=16\pi \sqrt{G}\sigma \bar A_0
\end{equation}
Notice that relation (\ref{S-Blaw}) resembles the 
Schuster-Blackett law, which is an empirical relation between
 the magnetic moments  and the angular momenta found in a wide
range of astrophysical objects from planets, to galaxies, including those
related to the presence of rotating neutron stars such as GRB 
or magnetars \cite{opher}. Let us mention that 
the observational evidence on this relation  
is still not conclusive. From observations, 
the $\beta$ parameter is found  to
be in the range 0.001 to 0.1. 

Imposing the PPN limits on the $\sigma$ parameter, we find
$\beta\lsim 10^{-4}$, which is just below the observed range.
Thus for a typical spiral galaxy, a direct calculation provides:
$B\sim \sigma 10^{-4}$ G, i.e. according to the PPN limits, the field
strength could reach $10^{-9}$ G without amplification. 

However, notice that even in the case in which this generation mechanism
took place, the determination of the 
actual amplitudes  of magnetic fields in astrophysical
objects would require to take into account the full 
magnetohydrodynamical evolution. Therefore, we do not expect 
the gravitationally generated magnetic field
to necessarily agree with observations. However this mechanism could 
help seeding standard amplification mechanisms such as dynamo 
with appropriate fields correlated to the object angular momentum.   


It is also interesting to evaluate the maximum magnetic
field that could be generated in  a Blackett-like experiment 
in laboratory \cite{nature}. Thus 
for a rotating neutral sphere of $M=500$ kg, radius $R=0.5$ m and
rotation frequency $\omega=100$ Hz, the field amplitude 
would be $B\sim \sigma 10^{-10}$ T
$\lsim 10^{-15}$ T, which is just below the fundamental sensitivity limit 
of SQUID or SERF magnetometers \cite{greenberg}.  

\section{Discussion}

Notice that in this scenario, it is the non-vanishing Ricci curvature
what generates electromagnetic fields in the presence of 
dark energy. Notice, however, that the only requirement for the main results of the present work is the presence of a cosmological electric potential. This implies that 
magnetic fields would be associated to the presence
of a non-vanishing energy-momentum distribution. 
In other words, the effect would be absent in vacuum even 
for curved backgrounds.  

Another interesting consequence of the presence of
 non-minimal couplings in the electromagnetic action (\ref{action})
is the fact that 
they play the role
of an effective mass term for the electromagnetic field during inflation.
This naturally provides an infrared cutoff in the calculation of the
field dispersion from quantum fluctuations \cite{Lyth}.

As shown before,  fields of  strengths up to $10^{-9}$ G 
could be generated on galactic scales in this theory, 
which 
could seed a galactic dynamo or even play the role of
"primordial" seeds and account for the observed 
magnetic fields in galaxies and clusters just by adiabatic 
compression in the collapse of the protogalactic cloud.
Therefore, a detailed study of the 
magnetohydrodynamical evolution in the presence of the 
gravitationally-induced current
will help establishing  
the  importance of dark energy  
in the origin of cosmic magnetic fields.   


\vspace{0.3cm}

{\em Acknowledgments:}
We would like to thank Prof. Misao Sasaki for useful comments. 
This work has been  supported by
MICINN (Spain) project numbers
FIS 2008-01323 and FPA
2008-00592, CAM/UCM 910309, 
MEC grant BES-2006-12059 and MICINN Consolider-Ingenio 
MULTIDARK CSD2009-00064. 
J.B. also wishes to thank support from the Norwegian Council 
under the YGGDRASIL project no 195761/V11.



\begin{thebibliography}{0}
\bibitem{Widrow}  L.~M.~Widrow,
  Rev.\ Mod.\ Phys.\  {\bf 74} (2002) 775; R.~M.~Kulsrud and E.~G.~Zweibel,
  Rept.\ Prog.\ Phys.\  {\bf 71} (2008) 0046091; P.~P.~Kronberg,
  Rept.\ Prog.\ Phys.\  {\bf 57} (1994) 325.
\bibitem{extragalactic} A. Neronov and I. Vovk, 
Science {\bf 328} (2010) 73; F.~Tavecchio, et al.,  
arXiv:1004.1329 [astro-ph.CO];  S.~Ando, A.~Kusenko,
  Astrophys.\ J.\  {\bf 722 } (2010)  L39.
\bibitem{Harrison} E.R. Harrison, MNRAS {\bf 147} (1970) 279; 
Phys. Rev. Lett. {\bf 30} (1973) 18
\bibitem{Turner} M.S. Turner and L.M. Widrow, {\it Phys. Rev.} {\bf D37}:
2743, (1988); K.~Bamba and M.~Sasaki, 
  JCAP {\bf 0702} (2007) 030. R. Durrer, L. Hollestein and R. Kumar Jain, arXiv:1005.5322 [astro-ph.CO]
\bibitem{QLS}J.~M.~Quashnock, A.~Loeb and D.~N.~Spergel, 
Astrophys. J.  {\bf 344} (1989) L49; T. Vachaspati, Phys. Lett. {\bf B265}
(1991) 258
\bibitem{Mota} O.~Bertolami and D.~F.~Mota,
  Phys.\ Lett.\  B {\bf 455} (1999) 96
\bibitem{pert} A.~L.~Maroto, 
  Phys.\ Rev.\  D {\bf 64} (2001) 083006;   
S.~Matarrese, S.~Mollerach, A.~Notari and A.~Riotto,
   Phys.\ Rev.\  D {\bf 71} (2005) 043502; K.~Ichiki, K.~Takahashi, H.~Ohno, 
H.~Hanayama and N.~Sugiyama, Science {\bf 311} (2006) 827;  
K.~Ichiki, K.~Takahashi, N.~Sugiyama, H.~Hanayama and H.~Ohno, 
 arXiv:astro-ph/0701329.
\bibitem{EM} J. Beltr\'an Jim\'enez and A.L. Maroto, 
JCAP {\bf 0903} (2009) 016; Phys.\ Lett.\  B {\bf 686} (2010) 175;
  Int.\ J.\ Mod.\ Phys.\  D {\bf 18} (2009) 2243;
 J.~B.~Jimenez, T.~S.~Koivisto, A.~L.~Maroto and D.~F.~Mota,
  JCAP {\bf 0910} (2009) 029
\bibitem{minimal} J. Beltr\'an Jim\'enez and A.L. Maroto,
arXiv:1010.3960 [astro-ph.CO] 
\bibitem{Will} C. Will, {\it Theory and experiment in gravitational physics},
Cambridge University Press, (1993)
\bibitem{VT}
  J.~Beltr\'an~Jimenez and A.~L.~Maroto,
  JCAP {\bf 0902} (2009) 025

\bibitem{HE} S.W. Hawking and G.F.R. Ellis, {\it The large scale structure of
space-time}, Cambridge (1973); G.~M.~Shore, arXiv:gr-qc/0302116.
\bibitem{Eling} C.~Eling, Phys.\ Rev.\  D {\bf 73} (2006) 084026
\bibitem{old} A. Schuster,   Proc. Lond. Phys. Soc. {\bf 24} (1912) 121;
A. Einstein,  Schw. Naturf. Ges. Verh. 105 Pt. 2, 85 (1924) 
S. Saunders S and H.R. Brown,  Philosophy of Vacuum
(Oxford: Clarendon) (1991); P. M. S. Blackett,   Nature {\bf 159} (1947) 658
\bibitem{early} W. Pauli, Ann. Phys. (Leipzig) {\bf 18}, 305 (1933);
J. G. Bennett et al., Proc. R. Soc. London A 198, 39
(1949); A. Papapetrou, Philos. Mag. {\bf 41}, 399 (1950);
G. Luchak, Can. J. Phys. {\bf 29}, 470 (1952); 
A. O. Barut and T. Gornitz, Found. Phys. {\bf 15}, 433 (1985).
\bibitem{limits} C. Amsler et al. (Particle Data Group), 
Phys.\ Lett.\  B {\bf 667} (2008) 1
\bibitem{Raffelt} G. Raffelt, Phys.\ Rev.\  D {\bf 50} (1994) 7729
\bibitem{opher}R. Opher and U.~F.~Wichoski,
  Phys.\ Rev.\ Lett.\  {\bf 78} (1997) 787;
R. da Silva de Souza, R. Opher, JCAP {\bf 1002} (2010) 022
\bibitem{nature} S.-P. Sirag, Nature 278, 535 (1979) 
\bibitem{greenberg} Ya. S. Greenberg, 
Rev.\ Mod.\ Phys.\  {\bf 70} (1998) 175
\bibitem{Lyth} D.~H.~Lyth, JCAP {\bf 0712} (2007) 016; 
  K.~Enqvist, S.~Nurmi, D.~Podolsky and G.~I.~Rigopoulos,
  JCAP {\bf 0804} (2008) 025; Y.~Urakawa and T.~Tanaka,
  Prog.\ Theor.\ Phys.\  {\bf 122} (2009) 779


 




 







\end{thebibliography}
\end{document}